\RequirePackage{ifpdf}
\ifpdf 
\documentclass[pdftex]{sigma}
\else
\documentclass{sigma}
\fi

 1 

\usepackage{amssymb}
\usepackage[mathscr]{euscript}
\usepackage{graphicx}
\usepackage[all,knot]{xy}
\usepackage{multicol} 

\newcommand{\iN}{\hbox{ {\leaders\hrule\hskip.2cm}{\vrule height .22cm} }}

\newcommand{\BL}{{\Big{[}}}
\newcommand{\BR}{{\Big{]}}}

\def\<{\langle} \def\>{\rangle}

\begin{document}

\allowdisplaybreaks

\renewcommand{\PaperNumber}{}


\ShortArticleName{Multisymplectic geometry and the notion of observables}

\ArticleName{Multisymplectic geometry \\ and the notion of  observables}

\Author{D. Vey\footnote{\footnotesize{email : dimitrivey@obspm.fr  \ mayaloop@gmail.com} } LUTH, Paris Observatory, Cosmology and Gravitation laboratory - University Paris 7}

\AuthorNameForHeading{D. Vey}


\Abstract{{\footnotesize{{\em This note  provides an overview of the notion of observable within the setting of multisymplectic geometry. We essentially follow the ideas described by F. H\'elein and J. Kouneiher  \cite{HK-01} \cite{HK-02} \cite{HK-03} and in particular in keeping  with the approach developed in \cite{H-02}. The following discussion should be considered as a synthesis of  the principal motives and results presented in those writings. This note is a contribution to the Frontiers of Fundamental Physics (FFP11) Conference Proceedings.
}}
}}



\section{Prolegomena}

\

The development of   {\em Quantum Gravity} $[{\bf QG}]$ is related to the construction of an adequate alphabet which would permit the building of  a bridge between the language  of differential geometry  (and its subdomain  riemaniann geometry) which forms the framework of Einstein's theory of {\em General Relativity} $[{\bf GR}]$,  and   the
  algebraic symbolism forming the framework of {\em Quantum Fields Theory} $[{\bf QFT}]$. In pursuing this aim,  we are led to consider the status of the following fundamental notions : {\em spacetime}, {\em observables}, {\em background structure} and {\em matter fields}.
Indeed, since we understand  {\em General Covariance} as the abandonment of any preferred
coordinate system for field equations for space-time  we  are led to consider the idea of  {\em background independence} as fundamental. For detailed discussion about the meaning of background independence and general covariance we follow the point of view  of J. Stachel \cite{stachel02}. This insight leads to the problem of observables in $[{\bf GR}]$ : for pure gravity {\em no} observables are given. This implies the rejection as meaningless of the notion of any {\em a priori} given {\em spacetime}  structure.  In such a picture the conclusion naturally follows that the  {\bf Relativity principle}\footnote{In this article we discuss Relativity Principles in the broad context of their development  from Galilean Relativity to Special
and General Relativiy.} is intrinsically rooted in the relativity of  {\em observable}. On this view of the philosophy of  $[{\bf GR}]$, the theory claims that observable quantities are not measured directly, but  are only compared to one another.  The key
idea of the multisymplectic approach is to give a precise definition of the notion of an observable and a method
to compare two observables without specify any volume form (which means for us without making any reference to a preferred spacetime background). In this approach, we recover a crucial insight : {\em dynamics} just tells us how to compare two observations. To
emphasize this fundamental point, we cite C. Rovelli \cite{CR-06} : {\em ''What has physical meaning is only the relative localization of the dynamical objects of the theory (the gravitational field among them) with respect to one another.''}

In canonical quantum gravity, one formulates the concept of {\em observable} on the ground of  Dirac \cite{Dirac}. The {\em Loop Quantum Gravity} $[{\bf LQG}]$ program \cite{ash} makes intensive use of Dirac observables. Dynamics is perceived as a gauge generated by first class constraints and an observable defined on the phase space commutes   with all the constraints. Drawing inspiration from and studying the intersection of the formalisms of Yang-Mills theory (dynamical connections) and $[{\bf QFT}]$ C. Rovelli and L. Smolin \cite{CR-07} made use of Wilson loop\footnote{The canonical variables of quantum geometry as developed in $[\bf LQG]$ are Wilson loops (given by the trace of the holonomy around the loop $\gamma$, $W_{\gamma} [{\cal A}] = \hbox{tr} [ h_{\gamma} [{\cal A}] ] $ of an SU(2)-connection ${\cal A}$) and fluxes of the conjugate momenta. Hence, $W [{\cal A}] $ is a functional of the connection that provides a rule for the parallel transport of  the SU(2) connection.} 
in $[{\bf QG}]$ context (since the functional on the space of connections  is invariant under gauge transformations) and studied the {\em loop representation} for Ashtekar variables. This brings to light the basis of Quantum Riemaniann Geometry. $[{\bf LQG}]$  is built upon two main kinematical operators  : the volume and the area operators. However neither of these is a Dirac observable, and neither commutes with all the canonical constraints. Hence, addressing this question of the  {\em physical}  meaning of such operators (even within kinematical area) highlights subtle issues concerning the question of an observable.
As this example emphasizes, an obstruction to establishing a
good notion of observable may be related to : $(\mathfrak{i})$ The issue of the set of Dirac first class constraints and
the subsequent notion of Dirac observable over phase space   
${\pmb{\cal O}}_{\tiny{\hbox{Dirac}}}$ (see {\bf 2.5}).
$(\mathfrak{ii})$ It seems that
suitable matter degrees of freedom need to be included to obtain a complete picture. $(\mathfrak{iii})$ The ever present issue of an observable {\em within} covariant field theory. Points $(\mathfrak{i})$ and $(\mathfrak{ii})$ exhibit  the necessity of Lepage-Dedecker theory for covariant Hamiltonian  field   theory. There, the Dirac constraint set can always be taken to be empty, this is possible by observing a total democracy between spacetime {\em and} matter fields. The last ($\mathfrak{iii})$ is much more subtle but forces us to recognize that any observable
quantity should emerge from intrinsic properties, namely from {\em dynamics}. In the
interplay between observables and dynamical considerations we recognize the astonishing beauty of the
{\em universal Hamiltonian formalism}.
  
 The purpose of this note is to illustrate the notion of  {\em observable} and the subtlety of this
concept  in multisymplectic geometry. Along the way we form an appreciation for the connection between a suitable notion of observable and the Relativity principle. Below, we
summarize four mathematical approaches to classical (and quantum) fields theory. Points {\bf 1.}$\mathfrak{1})$ and  {\bf 1.}$\mathfrak{2})$ are cornerstones of covariant canonical quantization beyond $[{\bf QG}]$.

\

$\mathfrak{1})$  {\bf  Multisymplectic Geometry}. [{\bf MG}] Within the context of covariant canonical
quantization [{\bf MG}] is a generalization of  symplectic  geometry  for field theory. It allows us to construct a general framework for the calculus of variations with several variables. Historically [{\bf MG}] was developed in three distinct steps :

$\quad \bullet$ Its origins are connected with the names of C. Carath\'eodory \cite{cara} (1929), T. De Donder \cite{Donder} (1935) on one hand and H. Weyl \cite{Weyl} (1935) on the other. We make this distinction since the motivations involved were different  :   Carath\'eodory and later Weyl, were involved with the
generalization of the Hamilton-Jacobi equation to several variables and the line of development
stemming from their work is concerned with the solution of variational problems in the setting ofÊ
the action functional. 
On the other hand, E. Cartan \cite{cartan} recognized   the crucial importance of developing an  {\em invariant language} for differential geometry not dependent on local coordinates. ÊIn connection with this aim we recover the motivations of T. De Donder.
The two approaches merge in the so-called De Donder-Weyl theory based on the affine  multisymplectic manifold ${\cal M}_{\tiny{\hbox{DDW}}}$.

$\quad   \bullet$  As was noticed by T. Lepage \cite{Lepage} this theory is a singular case of a more general multisymplectic theory. P. Dedecker \cite{KS12} (1953) set out to map the path to its geometrization. Indeed, for field theory, we are led to think variational problems as  $n$-dimensional submanifolds $\Sigma^{n}$ embedded in a
$(n+k)$-dimensional manifold ${\cal N}^{n+k}$. One observes the key role of the {\em Grassmannian bundle} as the analogous of the tangent bundle for variational problems for fields theory. 

$\quad  \bullet$ Finally, a very active Polish school in the seventies  further developed the geometric setting.   W. Tulczyjew, J. Kijowski, K. Gawedski and W. Szczyrba formulated important steps. \cite{ KS1} \cite{JKWMT} \cite{GAGA}. We find already in their work the notion of an {\em algebraic form}, and in the work  of J. Kijowski \cite{JK-01} a corresponding formulation of the notion of   a  {\em dynamical observable} emerges.  

 \

$\quad \mathfrak{2})$ {\bf Covariant Phase Space} approach  [{\bf CPS}]  share many features with [{\bf{MG}}]. The main idea is that we are not working on ordinary phase space but rather on the space of solutions ${\cal E}^{\cal H}$ of a Hamiltonian\footnote{ It also may be perceived from a Lagrangian standpoint :   [{\bf CPS}]    of a lagrangian field theory is the solution space
of the associated Euler-Lagrange equations.}
dynamical problem, namely a functional
space. As noticed by G. J. Zuckerman \cite{Zuk} and also though the work of Goldschmidt-Sternberg \cite{HGSS} (1973) and Crnkovic-Witten \cite{Witten0} (1987) the key observation is to define a  canonical pre-symplectic structure\footnote{We follow the notation of F. H\'elein in \cite{H-02} and postpone  comments on the construction of the symplectic structure $\pmb{\varpi}= \delta\Theta^{\Sigma}$ defined on  [{\bf CPS}]  at remark $\mathfrak{5})$ of section {\bf 4}. } $\pmb{\varpi}= \delta\Theta^{\Sigma}$ on such functional space. We will not enter into details here but we refer to the ingenious paper of  F. H\'elein \cite{H-02} which draws (on  the work of Szczyrba-Kijowski (1976) \cite{KS1}, see also \cite{Romero}), the relation of [{\bf CPS}] approach with multisymplectic geometry in a modern view. One observes : 

$\quad   \bullet$  The same fundamental mathematical entity actually present in both approaches, [{\bf MG}] and [{\bf CPS}], clearly manifested in the {\em $n$-phase space} notion. 

$\quad   \bullet$  We should underline that the  {\em invariant language}  provided crucial insight and merge out in the theory of integral invariants (first performed by H. Poincar\'e and further developed by E. Cartan \cite{cartan}). Within this approach to dynamical principles (what we may call {\em Cartan principle of dynamics}) we find a deep connection between [{\bf MG}] and [{\bf CPS}]. This relation  may be seen as a modern continuation of T. De Donder attempt in seeking to extend the notion of integral invariants to variational problems with several variables.
 
 \
 
$\quad   \mathfrak{3})$  {\bf Algebraic geometry-topology} This approach\footnote{using  tools as ${\cal C}$-spectral sequence or variational bi-complex development of the Vinogradov secondary calculus today  centres mainly on the work of the so-called {\em diffiety school}.} is based on the work of Vinogradov (1984) and the so-called secondary calculus. \cite{Vin}
It addresses within a beautiful cohomological setting the idea of  a {\em local} functional differential
calculus on the space of solutions of a generic system of {\em partial differential equations}. And in doing so, emphasizes the strong relation with [{\bf CPS}], see \cite{Vit}.

\

$\quad   \mathfrak{4})$  {\bf Algebraic Quantum Field Theory} [{\bf aQFT}] (see \cite{BFV-01}) is connected to the study of {\em  formal functional methods} and is to be understood, from a physical standpoint as motivated by the needs
of [{\bf QFT}] (the path integral approach). A further question  concerns the possibility of formulating a consistent axiomatic Quantum Field Theory within arbitrary curved spacetimes.
 We mention
this approach in order to emphasize a crucial point about {\em causal structure} since [{\bf aQFT}]  is based upon two
main principles : {\em covariance} and {\em locality}. Indeed, in  [{\bf QFT}]  the causal structure is  fixed\footnote{Then one observes a well-defined causal structure for example for a scalar field $\pmb{\Phi} (x) $ with spacelike interval between the points $x$ and $x'$ we have causality relations : $ [  \pmb{\Phi} (x) ,  \pmb{\Phi} (x')  ]  = \pmb{\Phi} (x) \pmb{\Phi} (x') - \pmb{\Phi} (x') \pmb{\Phi} (x)    = 0 $.}  and we
have a preferred notion of causality  :   the existence of a non-dynamical, Minkowski background metric $\eta$. In $[{\bf GR}]$ the situation is drastically different. {\em Since no prior geometry is given, what is the meaning of such relation?} Our intuition is that it is the area where [{\bf MG}] and [{\bf CPS}] overlap, formally and conceptually, that we shall discover the necessary tools to resolve this very subtle question.
 
In the following, we concentrate on {\em multisymplectic approach} : classical field theory is   treated on a {\em finite dimensional} framework. Let us first examine the symplectic roots of dynamics.

\section{Hamiltonian system and Symplectic roots}

We cite Dirac \cite{Dirac}, who believed {\em "there will always be something missing which we can only get by working from Hamiltonian, or maybe from some {\em generalisation} of the concept of a Hamiltonian.''} Hamiltonian dynamics rests on {\em symplectic geometry} whose key ingredient is a closed, non degenerate differential 2-form. Intrinsic measurements on a symplectic manifold are then 2-dimensional. In addition,  current developments    indicate that symplectic geometry is strongly connected  to  complex geometry.\footnote{Symplectic geometry is the backdrop of the canonical approach but it also appears in areas such as {\em duality theory} and {\em string theory}. A famous avatar of this interaction is the mysterious mirror symmetries between symplectic manifolds.
J-holomorphic curves (a kind of symplectic geodesics) exhibit some connection with quantum cohomology.}

\subsection{From Lagrangian to Hamiltonian dynamics}

A simple case where symplectic geometry  appears in physics is illustrated by time-independent mechanics.
Let us consider $\mathfrak{Z}$ the configuration bundle and a Lagrangian density   defined on a set of smooth paths $\gamma : I \subset \Bbb{R} \longrightarrow {\mathfrak{Z}} $, then, from the Lagrangian density \eqref{ELEL1} $(\mathfrak{i})$ $L : T\mathfrak{Z} \longrightarrow \Bbb{R}$, we obtain the well known Euler-Lagrange equations \eqref{ELEL1}$(\mathfrak{ii})$.
 \begin{equation}\label{ELEL1}
 (\mathfrak{i})
 \quad
{\cal L} (\gamma) = \int_{I} L(\gamma(t),{\dot \gamma}(t)) dt 
\quad
\quad
\quad
\quad
 (\mathfrak{ii})
 \quad
{d\over dt}  \BL  \frac{\partial L}{\partial z^{i}}  (\gamma(t),{\dot \gamma}(t))  \BR  = \frac{\partial L}{\partial q^{i}} (\gamma(t),{\dot \gamma}(t)) 
\end{equation}
In the Lagrangian setting, we work with the tangent bundle  $T\mathfrak{Z}$.
Then we switch into the Hamiltonian setting  (and work with cotangent bundle $T^\star \mathfrak{Z}$) by performing a Legendre transform\footnote{We consider a Legendre transform, with no degeneracy, therefore we can define its inverse $\mathfrak{J}^{-1}$ (left side of equation\eqref{LTLT}). Let consider $(q, {\cal Z} (q,p) )= \mathfrak{J}^{-1} \BL (q,p) \BR $. The {\em Legendre transform hypothesis} gives us a a characterization of ${\cal Z} (q,p)$ : 
$
p_{i} (t) =  {\partial L } / {\partial z^{i}} (q, {\cal Z} (q,p)) $ and $ {\cal Z}^{i} (q,  {\partial L }/ { \partial z } (q,z)  ) = z^{i}
$.
} 
\begin{equation}\label{LTLT}
\begin{array}{cccl}
\mathfrak{J} : & T\mathfrak{Z} & \longrightarrow & T^\star\mathfrak{Z}\\
& (q,z) & \longmapsto & \left( q,{\partial L\over \partial z}(q,z)\right)
\end{array}
\quad \quad 
\quad
\begin{array}{cccl}
\mathfrak{J} ^{-1}: & T^\star\mathfrak{Z} & \longrightarrow & T\mathfrak{Z} \\
& (q,p) & \longmapsto &  \left( q,z\right) = \left( q,{\cal Z}(q,p)\right)
\end{array}
\end{equation}
One replaces  Euler-Lagrange system of equations involving $(\gamma(t), \dot\gamma(t))$ in a new system of equations involving\footnote{ We define $\pi_i (t)$ as $
\pi_i (t) = {\partial L}  / {\partial {{z^i}}} (t) =  {\partial L} / {\partial z^{i}} (\gamma(t), \dot \gamma (t))
$. Hamiltonian dynamics express time evolution of
coordinates $(q,p)$ on $T^{\star}\mathfrak{Z}$, when $(q,z)$ satisfy  Euler-Lagrange equations.} $(\gamma(t), \pi(t))$.  This is possible thanks to the Hamiltonian function : $H:    T^{\star} \mathfrak{Z} \longrightarrow   \Bbb{R} $ defined $\forall (q,p) \in T^{\star} \mathfrak{Z}$,  
$
 H(q,p) = p_i {\cal Z}^{i} (q,p) - L \circ  \mathfrak{J}^{-1}
$. Finally, one concludes that $\gamma : \Bbb{R} \longrightarrow \mathfrak{Z}$ is a solution of the Euler-Lagrange equations if and only if the map $\mathfrak{z}  : \Bbb{R} \longrightarrow  T^{\star}\mathfrak{Z} \ / \ t \longrightarrow \mathfrak{z}  (t) = (\gamma(t) , \pi(t) ) $ is a solution of the Hamilton equations :
\begin{equation}
\left\{
\begin{array}{ccc}
\displaystyle \frac{\partial {  H}}{\partial q^{i} } (\gamma(t),\pi(t))  =   - \frac{d {\pi}_{i}}{dt} (t)  &  \quad \quad \quad
& \displaystyle   \frac{\partial {  H} }{\partial p_{i} }  (\gamma(t),\pi(t))  =     \frac{d \gamma^{i} }{dt} (t)
\\
\end{array}
\right.
\end{equation}

Symplectic geometry is the natural arena for describing Hamiltonian dynamics. A symplectic
structure $\Omega$ defined on a manifold $\cal M$ is a closed ($\hbox{d}\Omega = 0$) and non-degenerate\footnote{The non degeneracy condition means that we can construct an isomorphism between the vector fields $X$ on $\cal M$ and the space of 1-forms : $T {\cal M} \rightarrow T^\star{\cal M} : \xi \rightarrow \xi \iN \Omega $. Then $\Omega$ is non degenrate means $\forall \xi \in T{\cal M}, \xi \iN \Omega  = 0  \Rightarrow \xi = 0$ } 2-form. The geometrization of Hamiltonian dynamics is performed via the use of canonical {\em Poincar\'e Cartan form} \eqref{coo}$ (\mathfrak{i})$ and the {\em symplectic 2-form} \eqref{coo}$ (\mathfrak{ii})$ given in coordinates :
\begin{equation}\label{coo}
 (\mathfrak{i})
 \quad
\theta = \sum_{1 \leq i \leq n} p_{i} dq^{i} \quad \quad \quad \quad 
 \quad \quad \quad \quad
 \quad
 (\mathfrak{ii})
 \quad
 \Omega = \hbox{d} \theta =  \sum_{1 \leq i \leq n} d p_{i} \wedge dq^{i}
\end{equation}
This process allows us to write the geometrical expression \eqref{geo1} for Hamiltonian mechanics given a Hamiltonian function $H :   T^{\star}\mathfrak{Z} \longrightarrow \Bbb{R}$  and a Hamiltonian vector field $\xi_H$.\footnote{with $\xi_H =  (\frac{\partial H}{\partial p_{i}}) \frac{\partial}{\partial q^{i} } - (\frac{\partial H}{\partial q^{i} }) \frac{\partial}{\partial p_{i} }$} 
\begin{equation}\label{geo1}
 \xi_H \iN \Omega = - \hbox{d}H 
\end{equation}
Integral curves of $\xi_H$ are formulated as maps
$\mathfrak{z} :
\left\{
\begin{array}{ccc}
\displaystyle  \Bbb{R} &  \longrightarrow
& \displaystyle    T^{\star}{\mathfrak{Z}} 
\\
\displaystyle t &
\longrightarrow & \displaystyle  \mathfrak{z} (t) =  (q(t) , p(t))  \end{array}
\right.$ that are solutions of  Hamilton's equations - they are the dynamical trajectories of the Hamiltonian system $({\cal M}, \Omega, H)$.
Namely we want to characterize the flow on $T^\star{\mathfrak{Z}}$ that is encoded in Hamilton equations.  
In fact, $\xi_H(q(t),p(t)) \in T(T^\star{\mathfrak{Z}})$ and a point $   \mathfrak{z}(t) =(q(t),p(t)) \in T^{\star} \mathfrak{Z}$  satisfying \eqref{1} describes it.
\begin{equation} \label{1}
{\hbox{d}  \over {\hbox{d} t} }(q(t),p(t)) = { {\hbox{d} \mathfrak{z}} \over {\hbox{d} t} } (t)  =  \xi_H (\mathfrak{z}(t)) =\xi_H (q(t),p(t))
\end{equation}
Hamilton's equations are  described by $\eqref{1}$, the map $\mathfrak{z} (t) $ parametrizes an integral curve of $\xi_H $.

\subsection{Hamiltonian constraint and presymplectic structure}

In the relativistic formulation based on the {\em extended configuration space} (namely  the set of points $(q^{\circ}(\tau) = \tau , q^{i}) \in \Bbb{R} \times {\mathfrak{Z}}  = \mathfrak{Z}^{\circ}$) and the {\em extended phase space} $T^\star{\mathfrak{Z}}^\circ$, we built a  geometrical picture analogous to the one developed for independent time mechanics, {\em e.g} to find a relation similar to  \eqref{geo1} but with objects (the Hamiltonian function and the symplectic 2-form) defined on $T^\star{\mathfrak{Z}}^\circ$. It leads to the relation \eqref{presy1} : 
\begin{equation}\label{presy1}
\xi_{\cal H} \iN \Omega = -\hbox{d}{\cal H}
\end{equation}
However \eqref{presy1} makes use of  the covariant Hamiltonian ${\cal H}( \tau ,q^{i}, p_{\circ} , p_{i}) = p_{\circ} + H(\tau,q^i,p_i) $  on $T^\star{\mathfrak{Z}}^\circ$\footnote{Since the space $T^\star \mathfrak{Z}^{\circ}$ is a cotangent bundle, then it carries a  canonical one-form ${\theta}= p_i dq^i + p_{\circ} d q^{\circ}   $ and symplectic form $\Omega = \hbox{d} \theta$. Let notice that if $dim(\mathfrak{Z}) = n$, then $dim(T^{\star}\mathfrak{Z}^\circ) = 2n + 2$. }. We exhibit pre-symplectic dynamics by introducing the constraint hypersurface \eqref{presy} ${\pmb{\Sigma}}_\circ$ as a $2n+1$ dimensional submanifold of $T^\star \mathfrak{Z}^{\circ}$ : 
\begin{equation}\label{presy}
{\pmb{\Sigma}}_\circ \subset T^\star({\mathfrak{Z}}\times \Bbb{R}) = {\Big{\{}} (q^{\circ},  q^{i}   ,p_{\circ} , p_{i}) \in T^\star({\mathfrak{Z}}\times \Bbb{R}) \quad /  \quad p_{\circ} = -H(q,p) {\Big{\}}} 
\end{equation}
Let  ${\mathfrak{i}}: {\pmb{\Sigma}}_\circ \rightarrow T^\star({\mathfrak{Z}}\times \Bbb{R})$    be inclusion map. The restrictions on the hypersurface ${\pmb{\Sigma}}_\circ$ : $  {\theta}_{|{{\pmb{\Sigma}}_\circ}} = {\mathfrak{i}}^*\theta $ and  $  {\Omega}_{|{{\pmb{\Sigma}}_\circ}} = {\mathfrak{i}}^*\Omega$  indicate a degenerate feature. This is the reason    ${\mathfrak i}^\ast{\theta}$ is called a {\em pre-symplectic} 2-form. (may be degenerate). Relativistic dynamic is given by the data of pre-symplectic space $({{\pmb{\Sigma}}_\circ} ,  {\Omega}_{|{{{\pmb{\Sigma}}_\circ}}})$ (with the additional condition that $ (\hbox{d} q^{\circ} )_{|{{{\pmb{\Sigma}}_\circ}}} \neq 0 $) whereas the pair $( T^\star{\mathfrak{Z}^{\circ}} , {\Omega})$ is symplectic. The analogous of \eqref{presy1}, namely dynamical equations, are now given by \eqref{presy3} :
 \begin{equation}\label{presy3}
 \forall \xi \in \Gamma (\mathfrak{Z}^{\circ}, T \mathfrak{Z}^{\circ}), \quad \quad \quad \quad (\xi \iN \Omega  ) |_{{\pmb{\Sigma}}_\circ}  = 0 \quad \quad \quad \quad \hbox{and} \quad \quad \quad \quad (\hbox{d} q^{\circ} )|_{{\pmb{\Sigma}}_\circ} \neq 0
 \end{equation}
 We come back later to the general case of n-phase space in which the pre-symplectic dynamics (of a Relativistic system) will be fully described  - see $\bf{3.4}$.$\mathfrak{2})$ and $\bf{3.4}$.$\mathfrak{3})$.

\subsection{Observable, dynamics and Poisson structure}

Thanks to Hamiltonian dynamics, we get a nice geometrical picture for the evolution of an
observable on the phase space $f : T^\star {\mathfrak Z} \rightarrow \Bbb{R} $. One either considers relation \eqref{didi}$(\mathfrak{i})$ which emphasizes the
  {\em symmetry}  based point of view. The evolution of an observable $\hbox{d} f$ is given though the means of the symplectic form and the related  vector field $\xi_f$ (see connection with Noether theorem in ${\bf 2.4}$). Or alternatively, one stresses  {\em dynamical}  insight with the relation \eqref{didi}$(\mathfrak{ii})$ : here evolution of an observable $\hbox{d} f / \hbox{d} t$ is given by Poisson Bracket with the Hamiltonian.
\begin{equation}\label{didi}
 (\mathfrak{i}) \quad \xi_f \iN \Omega = - \hbox{d}f \quad  \quad \quad \quad \quad 
 \quad  \quad
 \quad  \quad \quad \quad \quad (\mathfrak{ii}) \quad \{ H , f \} = \frac{\hbox{d}f}{\hbox{d}t}   
 \end{equation}
We are naturally led to consider the {\em Poisson Bracket} between observables \eqref{aa1}. 
\begin{equation}\label{aa1}
 \{ f , g \} = \Omega(\xi_f , \xi_g) = \xi_f \wedge \xi_g \iN \Omega 
\end{equation}
Indeed, we have $ \{ f , g \} = \xi_f (g) = \hbox{d} g (\xi_f) = {\mathfrak{L}}_{\xi_f} g = - {\mathfrak{L}}_{\xi_g} f $ and 
$\Omega (\xi_f , \xi_g)  =  \xi_g \iN ( \xi_f \iN \Omega) = - \xi_g \iN \hbox{d} f = - \hbox{d} f (\xi_g)  = - \{ g , f \}  = \{ f , g \}  $.
Notice that  agreement with \eqref{aa1} leads in  coordinates to the expression $\{ f , g \} = \sum_{i}{\big{(}} \frac{ \partial f}{ \partial {p_{i}}} \frac{ \partial g}{ \partial {q^{i} } } - \frac{ \partial f}{ \partial {q^{i}} } \frac{ \partial g}{ \partial {p_{i}} }{\big{)}}$. Thus we foand justify good ordering within the bracket in \eqref{didi}$(\mathfrak{ii})$. In the spirit of the canonical
approach, the formulation of Hamiltonian dynamics is a good preliminary for development of the
quantum theory. Indeed, if we replace the functions in $T^\star {\mathfrak{Z}}$
by Hermitian self-adjoint operators
and the Poisson bracket by the commutator  ${\big{[}} ; {\big{]}}$ roughly speaking  we are led to the {\em Heisenberg picture} that is contained in \eqref{Hei}, the quantum evolution equation.
\begin{equation} \label{Hei}
{\Big{[}}  \widehat{H}  ;  \widehat{F} {\Big{]}}  = i \hbar \ \frac{\hbox{d} \widehat{F}}{\hbox{d}t}
\end{equation}

\subsection{Hamiltonian symmetry and Noether theorem}

$\quad \mathfrak{1})$ {\bf Symplectomorphisms and Hamiltonian vector fields} 

The fundamental symmetry of classical
mechanics is contained in one single statement : ${\mathfrak{L}}_{\xi} \Omega = 0$. The vector field\footnote{$ \mathfrak{X}({\cal M}) $ denotes the space of vector fields on ${\cal M}$ and $\mathfrak{L}$ is the Lie derivative.}  $\xi \in  \mathfrak{X}({\cal M}) $ that leaves the symplectic form invariant is called {\em symplectic} and is  a generator of infinitesimal canonical transformations. We denote by
 $\mathfrak{H}_{\tiny{\hbox{sym}}} ({\cal M})$ the set of symplectic vector fields. 
The picture is alternatively described by the {\em flow}\footnote{The {\em flow} is a one parameter family of diffeomorphisms. In this case, the infinitessimal generator of this parameter family is the vector field canonically defined as $\xi_f =  \frac{d}{dt} {\big{[}} f \circ \phi_t {\big{]}}_{t = 0} $}  $\phi_t^{\xi} : {\cal M} \rightarrow {\cal M}$ induced by $\xi$. Then, a vector field $\xi \in \mathfrak{H}_{\tiny{\hbox{sym}}} ({\cal M})$  if its flow is defined by a symplectomorphism $(\phi_t^ \xi)^* \Omega = \Omega$. A {\em (locally) Hamiltonian} vector field is a symplectic vector field $\xi \in \mathfrak{H}_{\tiny{\hbox{sym}}} ({\cal M})$ that locally verify $\xi = \xi_f$ : it exists $f \in C^{\infty}({\cal M})$ such that $\xi_f \iN \Omega = -\hbox{d}f$. We denote by $\mathfrak{H}_{\tiny{\hbox{loc}}} ({\cal M})$ the set of Hamiltonian vector fields.

Now we are interested in continuous symmetry for a Hamiltonian system $({\cal M}, \Omega, H)$ composed of a symplectic manifold   $({\cal M}, \Omega)$ and a Hamiltonian function $H :{\cal M} \rightarrow \Bbb{R} $.  Such a symmetry is given by a vector  field  which preserves both the Hamiltonian function ({\em e.g} ${\cal L}_ \xi H = 0 $), {\em and} the symplectic form.  (We demand that $\xi \in \mathfrak{H}_{\tiny{\hbox{loc}}} ({\cal M}) $.)\footnote{We notice that we can formulate the interplay of  $\mathfrak{H}_{\tiny{\hbox{sym}}} ({\cal M})$ and $\mathfrak{H}_{\tiny{\hbox{loc}}} ({\cal M})$ from  a cohomological standpoint. Indeed,  if $\xi \in \mathfrak{H}_{\tiny{\hbox{sym}}} ({\cal M})$ then $\xi \iN \Omega$ is closed {\em e.g} $\hbox{d} (\xi \iN \Omega) = 0 $ whereas if  $\xi \in \mathfrak{H}_{\tiny{\hbox{loc}}} ({\cal M})$ then $\xi \iN \Omega$ is exact {\em e.g} it exists a function $f : {\cal M} \longrightarrow \Bbb{R}$ such that $\xi_f \iN \Omega = \hbox{d} f$. We observe the following exact sequence
$
 0 \longrightarrow \mathfrak{H}_{\tiny{\hbox{loc}}} ({\cal M}) \longrightarrow \mathfrak{H}_{\tiny{\hbox{sym}}} ({\cal M}) \longrightarrow H^{1}_{\tiny{\hbox{deRham}}}({\cal M}, \Bbb{R})\longrightarrow 0 
 $. Here $H^{1}_{\tiny{\hbox{deRham}}}({\cal M} ) $ the first de Rham cohomology (if  $H^{1}_{\tiny{\hbox{deRham}}}({\cal M} ) $ is trivial, then any symplectic vector field is Hamiltonian, globally.) measures the obstruction for a symplectic vector field to be Hamiltonian. The sequence is actually an exact sequence of
Lie algebras.} We conclude that :{\em ''Locally, any Hamiltonian vector field $\xi \in \mathfrak{H}_{\tiny{\hbox{loc}}} ({\cal M})$ is an infinitesimal generator of symplectomorphism"}
\

$  \mathfrak{2})$ {\bf Noether theorem}. 

The Noether theorem is fundamental in physics. It allows us to
derive either a relation between global symmetries and conserved charges or a relation between local
symmetries and gauge identities. Roughly speaking, from the Lagrangian standpoint, the Noether
theorem makes a connection between continuous families of symmetries of Lagrangian systems
and their first integrals. It states that : {\em ''Any differentiable smooth symmetry of the action
of a physical system has a corresponding conservation law.''}

Within the Hamiltonian picture, it is easy to exhibit the Noether theorem.   Let $\xi \in   \mathfrak{H}_{\tiny{\hbox{sym}}} ({\cal M})$ be a symplectic vector field, since ${\mathfrak L}_\xi \Omega = \hbox{d} (\xi \iN \Omega) + \xi \iN \hbox{d} \Omega  = 0$ and $\Omega$ is closed we get  ${\mathfrak L}_{\xi} \Omega = \hbox{d} (\xi \iN \Omega) = 0$. Thanks to the Poincar\'e theorem,  $\xi \iN \Omega$ is {\em locally} exact, so there exists a function $f : {\cal M} \rightarrow \Bbb{R}$ such that $\xi \iN \Omega = \hbox{d} f$, thus  $\xi \in   \mathfrak{H}_{\tiny{\hbox{loc}}} ({\cal M})$. Again in other words : if the Hamiltonian system 
 $ {\big{(}} {\cal M}, \Omega, H  {\big{)}} $ has a symmetry  $\xi$, then locally we have $\xi=\xi_f$ and $f$ is a constant of motion. On the other hand, if a function $f$ over the phase space  is constant, then $\xi=\xi_f$ such that $\xi \iN \omega = \hbox{d}f$ is a symmetry of the system. 
 
Finally, an obvious consequence is obtained from Lagrangian symmetry $\xi_f$ directly by consideration of  the Poincar\'e-Cartan canonical form. Hence, in this case $ {\mathfrak{L}}_{\xi_f} \theta$ is exact. Namely for any observable $f \in C^{\infty}({\cal M})$ we have 
 $
  {\mathfrak{L}}_{\xi_f} \theta = \hbox{d} ( \xi_f \iN \theta ) + \xi_f \iN \hbox{d} \theta = \hbox{d}  {\big{[}}   \xi_f \iN \theta  - f  {\big{]}}  
$. 
If we denote $\Gamma \subset T^{\star} {\cal M}$ a Hamiltonian curve, then a conservation law is expressed as $ \hbox{d} f |_\Gamma = 0$.

\subsection{Dirac constraints and Dirac observables}

We will not enter in the details of the so-call Dirac constraints and the related Dirac-Bergman
canonical quantization program \cite{Dirac} \cite{HT}.  However we give an indication of its treatment of the notion of an observable. Gauge invariance leads to a degenerate Legendre transform (canonical variables are related though {\em Legendre constraints}). The treatment of these constraints in the context of Hamiltonian dynamics leads\footnote{In particular we find the notion of first class and second class constraints.} to the notion of the constraint submanifold $\pmb{\Sigma} \subset T^{\star} {\cal M}$ of the phase space of the theory. $\pmb{\Sigma}$ is defined by the data of $m$ constraints $ \{ {\pmb{\chi}}_{i} (q,p) \}_{ 1 \leq i \leq m} = 0$.
In the context of Hamiltonian constrained theory with first class constraint algebra, a Dirac observable ${\pmb{\cal O}}_{\tiny{\hbox{Dirac}}}$ is a function over $T^{\star} {\cal M}$ such that\footnote{A Dirac observable ${\pmb{\cal O}}_{\tiny{\hbox{Dirac}}} $ have weakly (on the constraint surface ${\pmb{\Sigma}}$)
vanishing Poisson brackets with all of the first class constraints.}$ {\big{\{}}  {\pmb{\cal O}}_{\tiny{\hbox{Dirac}}} (q,p) ,  {\pmb{\chi}}_{i}    {\big{\}}} \thickapprox 0  $. A Dirac observable ${\pmb{\cal O}}_{\tiny{\hbox{Dirac}}} $ is defined as a physical, gauge-invariant quantity. Then dynamics with respect to first class constraints ${\pmb{\chi}}_{i}$ is perceived as a gauge.  We picture evolution of a function ${ {\cal F}} (q,p) $ over phase space \eqref{Dirac0}$(\mathfrak{i})$.
\begin{equation}\label{Dirac0}
(\mathfrak{i})
\quad
\frac{\hbox{d}}{\hbox{d} t} {\big{(}}  { {\cal F}} (q,p) {\big{)}} = {\big{\{}} {\cal H}  ,   { {\cal F}} (q,p) {\big{\}}}   
\quad 
\quad 
\quad 
\quad 
\quad 
\quad 
\quad 
\quad 
(\mathfrak{ii}) 
\quad
 {\big{\{}} {\cal H}  , {\pmb{\cal O}}_{\tiny{\hbox{Dirac}}} {\big{\}}}  \thickapprox 0
\end{equation}
In generally covariant systems like $[{\bf GR}]$,  the covariant Hamiltonian ${\cal H}  $ (which generates dynamics thus {\em time} evolution)  is a constraint which vanishes identically (as a sum of first class constraints). A Dirac observable ${\pmb{\cal O}}_{\tiny{\hbox{Dirac}}}$ is given   \eqref{Dirac0}$(\mathfrak{ii})$. It is this feature which gives rise to the problem of time in $[{\bf QG}]$. This is also a reflection of the interplay between reparametrization invariance and the schizophrenic status of time : seen as a dimension or parametrization variable. 

In the context of a universal Hamiltonian formalism it is always possible to work in a more
general Lepagean equivalent theory. In this case, thanks to the introduction of (huge) unphysical
variables one can subsequently make disappear first class Dirac constraints set. This point should underline the necessity of studying  gauge theory upon the basis of Lepage-Dedecker $[{\bf LD}]$ theory.

\section{Multisymplectic geometry}

\subsection{Toward Covariant Hamiltonian Fields Theories}

In a finite dimensional classical system, the motivation for choosing the cotangent bundle as a
mathematical model for phase space lay in the possibility of identifying elements of $T^{\star}{\cal M}$ with initial data for the dynamical evolution. 
Analogously, in a field theory we would expect the state space to consist of all Cauchy data for the system under  consideration. In the canonical approach to a  standard field theory, canonical variables are defined on spacelike hypersurfaces. 
Indeed, all the points on such a surface are at  {\it{equal time}} and the dynamical equations specify
how the canonical variables evolve from one equal time hypersurface to another. Therefore :

$\bullet$ We have an instantaneous Hamiltonian formalism on a infinite dimensional phase space.

$\bullet$ Space and time are  treated {\em asymetrically}, and thus we have a {\em non covariance scheme}.

In addition to the difficulties encountered in the classical  Hamiltonian regime when treating the field theory canonically, we have others when we quantify the theory. Stone-Von Neuman theorem does not apply to the infinite  dimensional case.\footnote{Then, there will be a large number of unitarily inequivalent representations of the  canonical commutations relations corresponding to inequivalent choices of the measure.} Moreover, in a large number of physically interesting cases in which the classical configuration is not a {\em linear vector space}, the question of the {\em measure} on the Hilbert space become intractable.\footnote{It leads at once to the problem of what could be the analogue of a {\it{distribution}} for such a non linear space?} The previous examples show something of the huge advantage gained by working on a  {\em finite} dimensional  formulation of canonical field theory, that allow to treat {\em space} and {\em time} in equal footing. This is precisely the insight of $[{\bf MG}]$.

The project  developed in  \cite{HK-01} \cite{HK-02} \cite{HK-03} concerns the construction
of a {\it{universal}} Hamiltonian formalism which generalize all the schemes of a manifestly covariant
 finite dimensional field theory, (\cite{Gotay} and references therein). This explains the adjective {\it{universal}}. The main focus in this construction is on the role of {\em Legendre correspondance}, and the hypothesis concerning the generalized Legendre condition. A motivation for the study of the universal Hamiltonian formalism is to apply it in the context of integrable systems and to analyse the canonical structure of physical theories  with the aim to quantify those theories. A key idea here is to construct a Hamiltonian description of classical field theory compatible with the principles of $[{\bf GR}]$.  Any effort towards understanding gravitation leads to the conclusion that spacetime should {\em emerge from} the dynamics,  therefore we need a description which does not assume any space-time-field splitting {\em a priori}. Spacetime coordinates should instead
emerge out from an analysis of what are the observable quantities and from the dynamics.

The framework of covariant Hamiltonian formalism for the calculus of variations with several variables addresses  the question : for general variational problems, is there an analogue of the
Hamiltonian theory for fields in 
$
{ \pmb{\Psi} }_{\bf u}:= \{{\bf u}:{\cal X}\longrightarrow {\cal Y}\}
$ which are critical points\footnote{Here, related to this Lagrangian functional of maps $u$ we have $ {\cal X} $ a n-dimensional manifold (space-time) and $
\hbox{d} \mathfrak{y} = dx^0\wedge dx^1\wedge ...\wedge dx^{n-1} $ is  a volume form on ${\cal X}$ and ${\cal Y}$ is a k-dimensional manifold (fields). } of :
$
{\cal L}[{\bf{u}}]  := \int_{\cal X} L(x,{\bf{u}}(x), \hbox{d}{\bf{u}}(x))\hbox{d} \mathfrak{y}
$.
The idea of a geometrical setting is to be able to treat more
general variational problems: the study of
$n$-dimensional submanifolds chosen in $ \pmb{\Psi} := \{ \Sigma^n \subset {\cal N}^{n+k}\}$
which are critical points of 
$
{\cal L}(\Sigma^n) := \int_{\Sigma^n}L(q,T_q\Sigma^n) \hbox{d} \mathfrak{y}
$. 
Here $\hbox{d} \mathfrak{y}$ is a $n$-form on ${\cal N}^{n+k}$. Particular examples are when $\Sigma^n$ is the graph in
${\cal N}^{n+k} = {\cal X}\times {\cal Y}$ of some some map
${\bf u}:{\cal X}\longrightarrow {\cal Y}$ or a section of a bundle. In this context, the analogue of the tangent bundle in mechanics is the {\em Grassmannian bundle}
$Gr^n{\cal N}$ of oriented $n$-dimensional subspaces of tangent
spaces to ${\cal N}$.  The analogue of the cotangent bundle
in mechanics is $\Lambda^nT^\star{\cal N}$.\footnote{  $\hbox{dim}(\Lambda^nT^*{\cal N})  > \hbox{dim}(Gr^n{\cal N}) + 1$ unless $n=1$
(classical mechanics) or $k=1$ (submanifolds are hypersurfaces). It is this feature that reveal Lepagean theories and the study of Generalized Legendre correspondence} The geometrization of Lepage-Dedecker theory  \cite{KS12}  \cite{Romero} \cite{HK-02} leads us seriously to consider the possibility of assuming
even less structure. We posit a  {\em complete democracy} between {\em time, space} {\bf and} {\em
internal variables }: these are  the {\em revendications} of Kaluza-Klein
theory, supergravity and superstrings theory.
Then the distinction between these variables should be a consequence of dynamical
equations. Viewed from this perspective the main focus is on :

$\bullet$ {\bf Observables} one follows F. H\'elein and J. Kouneiher \cite{HK-02} \cite{HK-03} to emphasize the dual point of view concerning the notion of observable forms. (namely {\em symmetry} v.s {\em dynamics} - see ${\bf 4.}$$\mathfrak{1}$ and ${\bf 4.}$$\mathfrak{2}$)

$\bullet$ {\bf Poisson Bracket} in the case of $[{\bf MG}]$, the Poisson structure is defined on forms  living in $\Lambda^nT^\star{\cal N}$, so far we are concerned with the appropriate bracket on $(n-1)$-forms. In the case of an $(n-1)$-form, the generalization is rather straightforward \eqref{gigi0} if we follow the {\em symmetry} point of view. However one see that a careful formulation of the {\em Relativity Principle},  from the viewpoint of observable forms makes it appear the {\em primitive seed} which  conjugates dynamical behavior {\bf and} Poisson structure,  in particular on $(p-1)$-form (with $1 \leq p \leq n-1$). So we gain insight into the inherent structure of this approach. In particular the question of the appropriate generalization of the  {\em Poisson bracket} on forms of arbitrary degree.  We are then faced with the induced notion of   {\bf pseudobracket}  (which is connected to {\em dynamical} considerations \eqref{gigi}). Moreover, the formulation of an adequate Poisson bracket for such a form of any degree makes use of the concept of {\bf copolarisation}, the avatar of of the notion of intrinsic dynamical property.

\

In the next sections, we will see that what lie at the heart of an {\em intrinsically dynamical geometry} are the related notions of {\em algebraic
observables} and {\em observable forms} leading to a well established set of {\em dynamical observables.}
 
\subsection{From Lagrangian to Hamiltonian dynamics}

Starting from the Lagrangian side of the picture, the principal objects of study are the fields, which are seen as maps $\mathbf{u}: \cal{X}\longrightarrow \cal{Y}$ with ${ {dim}({\cal X}) = n}$ and ${{dim}({\cal Y}) = k}$. We recall that from the Lagrangian action \eqref{eulerlagrange0}$(\mathfrak{i})$, we obtain the {\em Euler-Lagrange system} of equations  \eqref{eulerlagrange0}$(\mathfrak{ii})$.
 \begin{equation}\label{eulerlagrange0}
 (\mathfrak{i})
 \quad
 \mathcal{L}[\mathbf{u}] = \int_{\cal{X}}L(x,\mathbf{u}(x),{\hbox{d}\mathbf{u}_x}) \hbox{d}{\mathfrak y} \quad \quad  \quad
  (\mathfrak{ii}) 
\quad \frac{\partial}{\partial x^\mu}\left(\frac{\partial L}{\partial v^i_\mu}(x,\mathbf{u}(x),{\hbox{d}\mathbf{u}_x})\right) = \frac{\partial L}{\partial y^i}(x,\mathbf{u}(x),{\hbox{d}\mathbf{u}_x})
\end{equation}
We pass to the Hamiltonian side thanks to the Hamiltonian function :
$
H(x^\mu,y^i, p^\mu_i) = p^\mu_i v^i_\mu - L(x^\mu,y^i, v^i_\mu)
$. Hamiltonian dynamics is obtained by a non degenerate Legendre transform. If $(x^\mu,y^i,{v^i_\mu})\longmapsto (x^\mu,y^i,{p^\mu_i})$ is a good change of variables, with $ {p^\mu_i}:= \frac{\partial L}{\partial v^i_\mu}(x^\mu,y^i,{v^i_\mu}) $, Euler-Lagrange equations \eqref{eulerlagrange0} are equivalent to  the \textbf{De Donder-Weyl} $[{\bf DDW}]$ system of equations:
\begin{equation}\label{Hamilton0}
 \frac{\partial \mathbf{u}^i}{\partial x^\mu}(x)  =  \displaystyle \frac{\partial H}{\partial p^\mu_i}(x^\mu,\mathbf{u}^i(x),\mathbf{p}^\mu_i(x))
  \quad \quad  \quad \quad   
\sum_\mu\frac{\partial \mathbf{p}^\mu_i}{\partial x^\mu}(x) = \displaystyle - \frac{\partial H}{\partial y^i}(x^\mu,\mathbf{u}^i(x),\mathbf{p}^\mu_i(x))
\end{equation}
A geometrization is given if we consider in $\Gamma \subset {\cal M} = \cal{X} \times \cal{Y} \times \hbox{End}({{\cal{Y}}^\star},{{\cal{X}}^\star})$ the $n$-dimensional submanifold such that :
$
\Gamma = {\Big{\{}}  (x^\mu,y^i,{p^\mu_i}) \ /  \  y^i = \mathbf{u}^i(x) \ , \ {p^\mu_i}=\frac{\partial L}{\partial v^i_\mu}(x^\mu,\mathbf{u}^i(x),{\partial_\mu\mathbf{u}^i}(x))   {\Big{\}}}
$. If we introduce on ${\cal M}$ the {\em pre-multisymplectic} $(n+1)$-form $
 \Omega = \hbox{d}p^\mu_i\wedge \hbox{d}y^i\wedge d\mathfrak{y}_\nu - \hbox{d} H \wedge \hbox{d}\mathfrak{y}
$, then $[{\bf DDW}]$ system of equations reads $ \forall \xi \in \Lambda^{n} T_{(x,{\bf u}(x), {\bf p}(x))}{\Gamma}$ under the form 
$ \xi \iN \Omega |_\Gamma= 0 \nonumber $.

\subsection{Multisymplectic manifolds : general framework}

The previous example give us an intuitive glimpse of the road ahead. Now we enter on to examination of the general  geometric framework beyond $[{\bf LD}]$ theory. We may alternatively consider maps $\mathbf{u}: {\cal{X}}^n\longrightarrow {\cal{Y}}^k$ or sections $  {\pmb{\Phi}}  :
{\cal{X}}  \longrightarrow {\mathfrak{Z} }$ of a vector bundle
$ \pi_\mathcal{X}: {\mathfrak{Z}^{n+k}}\longrightarrow \mathcal{X}^n.$ We consider first order Lagrangian density $L(x,{\pmb{\Phi}}, \hbox{d} {\pmb{\Phi}})$ variational problems.
Let $\Lambda^nT^\star\mathfrak{Z}$ be the vector bundle  of $n$-forms over $\mathfrak{Z}$. One defines canonically the {\em Poincar\'e-Cartan canonical} $n$-form $\theta$ \eqref{PC}.
on $\Lambda^nT^\star{\mathfrak{Z}}$ :
\begin{equation}\label{PC}
\forall z\in \mathfrak{Z} \quad \forall p\in \Lambda^nT^\star_z\mathfrak{Z}
\quad 
\quad
\quad 
\quad
\theta_{(z,p)}(X_1,\cdots,X_n):=
p(\hbox{d}\pi_\mathcal{X}(X_1),\cdots, \hbox{d}\pi_\mathcal{X}(X_n))
\end{equation}
Then, the {\em multisymplectic} $(n+1)$-form $\Omega$ is defined as the exterior derivative of the Poincar\'e-Cartan form. $ \Omega = \hbox{d}\theta$. In order to stress the analogy with the case of classical mechanics case, we give the expression of $\theta$ and $ \Omega $ in local coordinates. Let denote $(q^{\mu})_{\tiny{1 \leq \mu \leq n+k}}$ coordinates on $\mathfrak{Z}$ (then a basis of 
$\Lambda^nT^\star{\mathfrak{Z}}$ is the family $(dq^{\mu_1} \wedge ... \wedge dq^{\mu_n})_{\tiny{1 \leq {\mu}_1  < ... {\mu}_n < n+k}}$) and $p_{{\mu}_1 ... {\mu}_n}$ coordinates on $\Lambda^nT^\star{\mathfrak{Z}}$. We obtain the expression for the canonical  $n$-form \eqref{PCmulti}$(\mathfrak{i})$ and the pataplectic $(n+1)$-form \eqref{PCmulti}$(\mathfrak{ii})$ (a straightforward generalization of the symplectic form \eqref{coo}$(\mathfrak{ii})$).
\begin{equation}\label{PCmulti}
(\mathfrak{i})
\quad
\theta = \sum_{1 \leq {\mu}_1  < ... {\mu}_n < n+k} p_{{\mu}_1 ... {\mu}_n} dq^{\mu_1} \wedge ... \wedge dq^{\mu_n}  
\quad \quad 
(\mathfrak{ii})
\quad 
 \Omega =   \sum_{1 \leq {\mu}_1  < ... {\mu}_n < n+k} dp_{{\mu}_1 ... {\mu}_n} \wedge dq^{\mu_1} \wedge ... \wedge dq^{\mu_n}
\end{equation}
One defines the notion of an $n$-{\bf multisymplectic manifold} $({\cal M}, \Omega )$ as a  manifold ${\cal M}$ endowed with an $(n+1)$-form $\Omega$ such that 
$\Omega$ is {\em closed} ($\hbox{d}\Omega = 0$) and  {\em non degenerate}. (for any vector $\xi$, if $\xi \iN \Omega = 0$, then
$\xi=0$). Then a {\em Hamiltonian function} on $(\mathcal{M},\Omega)$ is a function
$\mathcal{H}:\mathcal{M}\longrightarrow \mathbb{R}$ such that $\hbox{d}\mathcal{H}\neq 0$. In this case we generally describe ${\cal H} (x,   {\pmb{\Phi}}  (x) ,  {\pmb{\Pi}}^{\Phi}  (x))$. With $ {\pmb{\Pi}}^{\Phi} (x)= p(x)$ the set of multimomenta.
Now, let's recall that the necessary and sufficient conditions on the map $\mathfrak{z}  :  x \rightarrow (q(x), p(x)) : {\cal X} \rightarrow  {\cal M}$ to ensure  that this map is related to critical points $ {\pmb{\Phi}}  : {\cal X} \rightarrow {\mathfrak Z} $. We are looking for the conditions on $x \rightarrow (q(x), p(x))$ such that it exists fields $x \rightarrow  {\pmb{\Phi}}  (x)$  that verify  :
 
$(\mathfrak{i})$ $\forall x  \ (x, {\pmb{\Phi}} (x),  \hbox{d} {\pmb{\Phi}} (x)) \longleftrightarrow (q(x),p(x))$ (The generalized Legendre condition is realized) 

 $(\mathfrak{ii})$ Fields $x \rightarrow {\pmb{\Phi}} (x)$ are solutions of Euler Lagrange equations for the Lagrangian. Following \cite{HK-01}, 
these conditions are satisfied if and only if $(q(x),p(x)) $ verify\footnote{where  ${\cal I} \subset \Lambda^n T^\star {\cal M}$ is the subspace generated by the $dx^{\alpha}$ and the $n$-vector $X\in \Lambda^nT_{(q,p)}{\mathfrak Z}$ writes $X = X_1 \wedge ... \wedge X_n $ and $\forall \mu = 1 ... n $ we have 
$
 X_\mu =  {\partial (q(x),p(x)) /  \partial (x^1,\dots ,x^n)} $} in $\Lambda^n T^\star {\mathfrak Z}$
\begin{equation}\label{gogogo}
(-1)^nX \iN \Omega = \hbox{d}{\cal H}\quad \hbox{mod }{\cal I}.
\end{equation}
Equations \eqref{gogogo} are the {\em generalized Hamilton equations}. Usually, without loss of generality we work with {\em decomposable}\footnote{We define the set of decomposable $n$-vector by $D^{n}_{m}{\mathfrak{Z}} := {\big{\{}} X_1 \wedge \cdots \wedge X_n \in \Lambda^{n}T_{m}{\mathfrak{Z}} \ / \ X_1, \cdots , X_n  \in T_m {\mathfrak{Z}} {\big{\}}} $ and also we use the notation $[X]^{\cal H}_{m} := {\big{\{}} X \in  D^{n}_{m}{\mathfrak{Z}}  \ / \   X \iN \Omega = (-1)^{n} \hbox{d} {\cal H}_m {\big{\}}} $ }  $n$-vector $X \in \Lambda^n T^\star {\mathfrak Z}$ such that $\hbox{d}\mathfrak{y} (X) = 1$.  The geometrical entity that
in the classical case described a curve in the phase space (a Hamiltonian trajectory) is now
seen as a $n$-dimensional object. Thus arises the central notion of an   {\em Hamiltonian n-curve} (We denote ${\cal E}^{\cal H}$ the set of all such Hamiltonian curve.) :  a n-dimensional oriented submanifold $\Gamma \subset {\cal M}$ such that :  
 \begin{equation}\label{tyty}
 \forall m \in \Gamma, \quad \quad \exists X \in \Lambda^{n}T_m \Gamma \quad  \quad  X \iN \Omega_m = (-1)^{n} \hbox{d} {\cal H}_m
 \end{equation}

 \subsection{Some reflections on universal Hamiltonian formalism}
 
 Let us notice, among others, some points which underline the legacy of the universal Hamiltonian formalism.

\

$\quad \mathfrak{1})$ {\bf De Donder-Weyl multisymplectic theory}. The universal multisymplectic manifold $(\Lambda^nT^\star\mathfrak{Z}, \Omega)$ is very large. Usually multisymplectic theory is seen from the $[{\bf DDW}]$ standpoint which means that one restricts the theory to the affine multisymplectic submanifold   ${\cal M}_{\tiny{\hbox{DDW}}} \subset \Lambda^nT^\star\mathfrak{Z}$. We construct this as a constrained set of coordinates $(z,p)\in \Lambda^nT^\star\mathfrak{Z}$ obtained via the interior product of two vertical\footnote{We call a vertical vector field any $\xi \in T_z\mathfrak{Z}$ such that  $\hbox{d}\pi_\mathcal{X}(\xi ) = 0$, then we denote $T^{\bf V}\mathfrak{Z} $ the set of vertical vector field.} vector fields $ \xi, \chi \in T^{\bf V}\mathfrak{Z} $.
\begin{equation}
{\cal M}_{\tiny{\hbox{DDW}}}  \quad = \quad  {\Big{\{}} (z,p)\in \Lambda^nT^\star\mathfrak{Z} \quad / \quad  \forall  \xi, \chi \in T^{\bf V}\mathfrak{Z}  \quad \xi \wedge \chi \iN p = 0 {\Big{\}}}
\end{equation}
In such a context, set $\theta^{{\tiny{\hbox{DDW}}}}:= \theta |_{{\cal M}_{\tiny{\hbox{DDW}}}}$ the restriction of $\theta$ to  ${\cal M}_{\tiny{\hbox{DDW}}}$. Working on ${\cal M}_{\tiny{\hbox{DDW}}}$  is equivalent to taking into account all the constraints\footnote{We work in the framework of a fiber bundle $\mathfrak{Z}$ as configuration space. Coordinates on ${\cal X}$  are denoted by $\{ x^{\mu} \}_{1\leq\mu\leq n} $ and allow us to choose a volume $n$-form $\hbox{d}\mathfrak{y} = dx^{1} \wedge \cdots \wedge dx^{n}$ on ${\cal X}$. Coordinates on $\mathfrak{Z}$ are denoted $\{ z^{i} \}_{1\leq i\leq n} $  Also, let denote $\mathfrak{e} = p_{1\cdots n}$, $p_{1\cdots (\mu-1)i(\mu+1)\cdots n }$, $p^{\mu_1 \cdots \mu_2}_{i_{1} \cdots i_2} = p_{1 \cdots (\mu_1 - 1) i_1 (\mu_1 +1) \cdots  (\mu_2 - 1) i_2 (\mu_2 +1) \cdots n  }$  etc...  Also we use the notation  $\hbox{d} \mathfrak{y}^{i_1 \cdots i_p}_{\mu_1 \cdots \mu_p} = \hbox{d} z^{i_1} \wedge \cdots \wedge \hbox{d} z^{i_p} \wedge {\big{(}} \partial_{\mu_1} \wedge \cdots \wedge \partial_{\mu_p} \iN \hbox{d} \mathfrak{y} {\big{)}} $ as well as $ \hbox{d} \mathfrak{y}_{\mu} = \partial_\mu \iN \hbox{d} \mathfrak{y} $.
 }
 $p^{\mu_1 \cdots \mu_2}_{i_{1} \cdots i_2} = 0 $ for all $j > 1$. We obtain the corresponding multisymplectic $(n+1)$-form ${\Omega}^{\tiny{\hbox{DDW}}}$ \eqref{ZZZ}.
\begin{equation}\label{ZZZ}
{\Omega} = \hbox{d} \mathfrak{e} \wedge \hbox{d}\mathfrak{y} + \sum_{j=1}^{n} \sum_{\mu_1<\cdots<\mu_j} \sum_{i_1<\cdots<i_j} \hbox{d} p^{\mu_1 \cdots \mu_j }_{i_1 \cdots i_j} \wedge \hbox{d} \mathfrak{y}^{i_1 \cdots i_j}_{\mu_1 \cdots \mu_j}
\quad {\Rightarrow} \  {\cal M}_{\tiny{\hbox{DDW}}} \  {\Rightarrow} \quad
{\Omega}^{\tiny{\hbox{DDW}}} =  \hbox{d} \mathfrak{e} \wedge \hbox{d}\mathfrak{y} + \sum_{\mu} \sum_{i} \hbox{d} p^{\mu}_i \wedge \hbox{d}z^{i} \wedge \hbox{d} \mathfrak{y}_{\mu}
\end{equation}

$\quad \mathfrak{2})$ {\bf Pre-multisymplectic scenario}. We can allow the multisymplectic form to be
degenerate. We define a 
{\em pre-multisymplectic manifold} $(\mathcal{M},\Omega)$ 
such that the  $(n+1)$-form $\Omega$ is closed ($\hbox{d}\Omega = 0$). Now let us construct a 
{\em volume} $n$-form {$\hbox{d} {\mathfrak{y}}$ (an everywhere non-vanishing n-form)}. This indicates the right notion of an {\em n-phase space}  as the data  $(\mathcal{M},\Omega,{\hbox{d}\mathfrak{y}})$. Therefore we express {\em dynamics} on a {\em level set} of ${\cal H}$.\footnote{We can construct canonically a pre-$n$-multisymplectic manifold. $(\mathcal{M}^{\circ},\Omega|_{\mathcal{M}^\circ},{\hbox{d}\mathfrak{y}} = \tau \iN \Omega |_{\mathcal{M}^\circ} )$. Here the $
 \mathcal{M}^\circ:= \mathcal{H}^{-1}(0):= \{ (q,p) \in \mathcal{M}|\ \mathcal{H}(q,p) =0\}
$ and $\tau$ is a vector field s.t. $\hbox{d}\mathcal{H}(\tau) = 1$.
} In this way we clarify the framework
which connects relativistic dynamical systems and the treatment of  {\em Hamiltonian constraint}. Now, {\em dynamical equations}  become in the pre-multisymplectic case \eqref{ZZZZ}. 
\begin{equation}\label{ZZZZ}
{\forall \xi \in C^\infty ({\cal M}, T_m{\cal M})  ,\quad  \quad (\xi \iN \Omega) |_\Gamma = 0 \quad \quad \quad \quad \hbox{and}  \quad \quad \quad \quad {\hbox{d}\mathfrak{y}}|_\Gamma \neq 0}.
\end{equation}
In this case, $\Gamma$ is called a Hamiltonian $n$-curve. The variational principle then states that, for all $n$-dimensional submanifold $\Gamma \subset {\cal M}$ on which ${\hbox{d}\mathfrak{y}}|_\Gamma \neq 0$, $\Gamma$ is a critical point of the functional $ \mathcal{A}[\Gamma]:= \int_\Gamma\theta
$ if and only if $\Gamma$ is a Hamiltonian $n$-curve.

Notice that we understand the geometrization of the $[{\bf DDW}]$  system \eqref{Hamilton0} as the expression of a particular choice of $[{\bf LD}]$ theory : we work on the manifold ${\cal M}_{\tiny{\hbox{DDW}}}$ and given a Hamiltonian function $\mathcal{H}: {\cal M}_{\tiny{\hbox{DDW}}} \subset \Lambda^nT^\star\mathfrak{Z}\longrightarrow \mathbb{R}$ we restrict further to the submanifold $\mathcal{H} = 0$. In this case,   multisymplectic $(n+1)$-form ${\Omega}^{\tiny{\hbox{DDW}}}$ \eqref{ZZZ} is : ${\big{[}} {\Omega}^{\tiny{\hbox{DDW}}} {\big{]}}_{{\cal H} = 0} = \Omega = \hbox{d}p^\mu_i\wedge \hbox{d}y^i\wedge \hbox{d} \mathfrak{y}_\mu - \hbox{d}H\wedge \hbox{d}\mathfrak{y}.$
  
  \
  
  $\quad \mathfrak{3})$ {\bf Canonical variable $\mathfrak{e}$}. We denote $\mathfrak{e} = p_{1\cdots n} $ seen as a canonical variable conjugate to the volume form $\hbox{d} \mathfrak{y}$. 
  Notice that $\mathfrak{e}$ does not enter in to the Euler-Lagrange equation. We can always
write  ${\cal H} (q,\mathfrak{e},p) = \mathfrak{e} + H(q,p) $ and then work in a level set ${\cal H}^{-1}(0)$  choosing appropriate variable $\mathfrak{e}$. The relativistic case described in ${\bf 2.2}$  is described as a 1-phase space, thanks to Hamiltonian constraint. The constraint hypersurface $\pmb{\Sigma}_\circ$ is identified with the level set ${\cal H}^{-1} (0) = {\big{\{}}  (q, p) \in  {\cal M}  = T^\star{\mathfrak{Z}}^\circ  /  {\cal H} (q,p) = 0  {\big{\}}}$ (and $\mathfrak{e}$ identify with $p_\circ$ in ${\bf 2.2}$.) Then,  $ {\big{(}}   {\cal H}^{-1} (0) =  { {\pmb{\Sigma}}_\circ } ,  {\Omega}_{|{{ {\pmb{\Sigma}}_\circ }}} , (\hbox{d} q^{\circ} )_{|{{ {\pmb{\Sigma}}_\circ }}}  {\big{)}} $ is a 1-phase space. : $ (\hbox{d} q^{\circ} )_{|{\Gamma}}  $ does not vanish and $\Omega_{ |{\pmb{\Sigma}}_\circ }$ is a  closed 2 form.
 
 \
 
$\quad \mathfrak{4})$ {\bf Infinitesimal symplectomorphism}.
We may define an {\em infinitesimal symplectomorphism} of $({\cal M}, \Omega)$ to be a vector field $\Xi \in \Gamma({\cal M} , T{\cal M})$ such that 
$
{\mathfrak{L}}_{\Xi} \Omega = 0  
$.
Therefore, using the Cartan formula, we obtain : $
{\mathfrak{L}}_{\Xi} \Omega  = \hbox{d} (\Xi \iN \Omega) + \Xi \iN \hbox{d} \Omega = 0.
$
so that this relation is equivalent to
$
\hbox{d} (\Xi \iN \Omega) = 0
$. We denote $\mathfrak{sp}_\circ (\cal M)$ the set of all symplectomorphisms of ${\cal M}$. 
  
 
 
 \
 
$\quad \mathfrak{5})$ {\bf Generalized Legendre correspondence} 
We pass from a Legendre transform to the
generalized Legendre correspondence that may always be chosen as being non degenerate.  It correspond to the shift from $[{\bf DDW}]$  to  $[{\bf LD}]$ theories. This is needed for a fully even  handed treatment between space, time and fields.  In the traditional $[{\bf DDW}]$
 approach  in terms of {\em contact structure} and {\em jet bundles}, \cite{Gotay} Lagrangian density $L : J^{1} \mathfrak{Z} \longrightarrow \Bbb{R}$ is described  on the  first order jet bundle over $\mathfrak{Z}$.  However, this (poly)multisymplectic approach involves a triviality of the extended phase space as a bundle over spacetime. Therefore it contain a duality between two categories : {\em spacetime} {\bf and} {\em fields}. (induced  decomposition of forms and multivectors along with vertical and horizontal components.) \cite{Kana-01} \cite{Pau}. One comes up against  this point when we want to construct observables $(p-1)$-forms in the section {\bf 4.}$\mathfrak{3})$.
 
 \
 
$ \mathfrak{6})$ {\bf The Geometric standpoint}  leads us to recognize the central role of the notion of a {\em graph}. Indeed, Hamiltonian $n$-curve are seen as  {\em graphs} of the generalized Hamilton equations \eqref{tyty}. We give a glimpse of this : 
 the system \eqref{Hamilton0} can be expressed with the help of  geometrical condition on the graph $\Gamma = G(\bf{u}, {\bf p}) \subset \cal{X} \times \cal{Y} \times \hbox{End}({{\cal{Y}}^\star},{{\cal{X}}^\star}) $. This independence condition $\hbox{d} \mathfrak{y} |_\Gamma  = \mathfrak{i}^\ast |_\Gamma  \hbox{d} \mathfrak{y} \neq 0 $ ($\mathfrak{i} : \Gamma \longrightarrow {\cal M} $ the inclusion map) is necessary to describe Hamiltonian n-curve {\em locally} as the graph of some map $({\bf u},{\bf p})$. An Analogous condition is found in the general case with the  {\em Grassmanian bundle}.\footnote{In this case {\em Lagrangian density} is now defined as $L : Gr^{[d\mathfrak{y}]}  { \cal{M} } \longrightarrow \Bbb{R}$ where $Gr^{[d\mathfrak{y}]}$ is the fiber bundle over ${\cal M}$ whose fiber over $m \in {\cal M}$ is the set of all oriented $n$-dimensional vector subspaces $T$ of $T{\cal {M}} $ with the condition $\hbox{d} \mathfrak{y} |_T > 0 $.} It permits us to expand Hamilton equations without writing Euler-Lagrange equations.

\section{Observable forms in multisymplectic geometry}

In this section we introduce the set of algebraic observable forms $\mathfrak{P}_{\circ}^{n-1}$ and of observable forms $\mathfrak{P}^{n-1}$, and finally the set of dynamical observable functionals ${\pmb{\cal O}}^{\cal H}$.  

\

$\quad \mathfrak{1})$ {\bf Algebraic observable $(n-1)$-form  [AOF]}.
A $(n-1)$-form $F$ on $({\cal M}, \Omega)$ is called {\em algebraic observable $(n-1)$-form} if and only if there exists $\xi_F$ such that $\xi_F \iN \Omega + dF = 0 $. We denote $\mathfrak{P}_{\circ}^{n-1}$ the set of all algebraic observable $(n-1)$-forms.
This standpoint reflects the {\em symmetry} point of view. It is a natural analogous to the  question of the Poisson bracket for classical mechanics.  \eqref{aa1} Then, $\forall F, G \in \mathfrak{P}_{\circ}^{n-1}$
 we define the Poisson Bracket \eqref{gigi0} : 
\begin{equation}\label{gigi0}
{\big{\{}} F , G {\big{\}}} = \xi_F \wedge \xi_G \iN \Omega =  \xi_F   \iN \hbox{d} G = - \xi_G   \iN \hbox{d} F 
\end{equation}
Actually  ${\big{\{}} F , G {\big{\}}} \in  \mathfrak{P}_{\circ}^{n-1}$ and bracket \eqref{gigi0} satisfy Jacobi structure modulo an exact term :
\[
{\big{\{}}  F  ,  G {\big{\}}} + {\big{\{}}  G  ,  F {\big{\}}} = 0
\quad \quad \hbox{and} \quad \quad
{\big{\{}}  \{F , G \} H  {\big{\}}} + {\big{\{}}  \{ G , H \} F {\big{\}}} + {\big{\{}}  \{ H  , F \} G {\big{\}}} = \hbox{d} (\xi_{F} \wedge \xi_{G} \wedge \xi_{H} \iN \Omega)
\]
Any algebraic infinitesimal symplectomorphisms may always be written\footnote{ The vector field $\Xi \in \Gamma({\cal M} , T{\cal M})$, also below  we consider  $ \zeta:= \sum_\alpha  \zeta^\alpha(q){\partial \over \partial q^\alpha}$
is an arbitrary vector field on ${\mathfrak Z}$. } $\Xi = \chi + \bar{\zeta}$. \cite{HK-02} The notions of {\em generalized  position} $Q^{\xi} = \sum \xi_{\mu_1 \cdots \mu_{n-1}} dq^{{\mu_1}} \wedge \cdots \wedge dq^{{\mu_{n-1}}}$ (with $\chi \iN \Omega = - \hbox{d} Q^{\xi}$) and {\em generalized momenta}   $ P_{\zeta} = \zeta \iN \theta$ (with $\bar{\zeta} \iN \Omega = - \hbox{d} P_{\zeta}$) make us write any $ F \in \mathfrak{P}_{\circ}^{n-1}$ as $F = Q^{\xi} + P_\zeta$. If we note $\mathfrak{sp}_Q (\cal M)$ and $\mathfrak{sp}_P (\cal M)$ the set of symplectomorphisms of the form $\chi$ and $\bar{\zeta}$ respectively we have the structure : $ \mathfrak{sp}_\circ({\cal M}) =\mathfrak{sp}_P({\cal M})\ltimes \mathfrak{sp}_Q({\cal M})$. (see theorem in \cite{HK-02}).
 
 \
 
$\quad \mathfrak{2})$ {\bf  Observable $(n-1)$-form } {[\bf OF]}  \cite{HK-02} \cite{HK-03} gives the right generalization of the quantum
Heisenberg evolution equation.  Then we  construct the {\em pseudobracket}  in relation \eqref{gigi}$(\mathfrak{i})$
\begin{equation} \label{gigi}
(\mathfrak{i})
\quad
{\big{\{}}  {\cal H} , F  {\big{\}}} \ \hbox{d}\mathfrak{y}  {\big{|}}_\Gamma  \ =  \  \hbox{d}F {\big{|}}_\Gamma 
\quad
\quad 
\quad 
\quad
\quad 
\quad 
(\mathfrak{ii})
\quad
{\big{\{}}  {\cal H} ,  F  {\big{\}}} \  \hbox{d} G  {\big{|}}_\Gamma  =  {\big{\{}}  {\cal H} , G {\big{\}}}  \ \hbox{d} F  {\big{|}}_\Gamma 
 \end{equation}
 The fact that ${\cal H}$ is not an observable form is related to the notion of Poisson {\em pseudobracket} 
   $ {\big{\{}}  {\cal H} , F  {\big{\}}}   $. Its philosophy is grounded in the  {\em dynamical}  aspect. The idea beyond is that given a point $m \in {\cal M} $ and a Hamiltonian function ${\cal H}$ ({with $X(m) \in [X]^{\cal H}_{m}$ } see footnote $\mathfrak{[24]}$), $\langle X(m) , \hbox{d} F_{m} \rangle  $ should only depend  on $\hbox{d}{\cal H}_{m}$. Now, from  \eqref{gigi}$(\mathfrak{i})$ we find the relation  \eqref{gigi}$(\mathfrak{ii})$ for all $F , G \in \mathfrak{P}^{n-1}$ and $\Gamma$, a Hamiltonian $n$-curve.
We observe that in the equation  \eqref{gigi}$(\mathfrak{ii})$, no volume form $\hbox{d}\mathfrak{y}$ is singled out : dynamics just prescribes
how to compare two observations. This naturally encapsulates the {\bf Relativity Principle}. It also allows us to understand the relationship of {\em dynamics} and {\em observables}.

We denote $\mathfrak{P}^{n-1}$ the set of such observable form. The interplay of both close concepts  $\mathfrak{P}_{\circ}^{n-1}$ and  $\mathfrak{P}^{n-1}$  leads to the notion of a {\em pataplectic manifold} : a multisymplectic manifold $({\cal M},\Omega)$ where the set of observable $(n-1)$-forms coincides with the set of algebraic observable $(n-1)$-forms  : $\mathfrak{P}_{\circ}^{n-1} = \mathfrak{P}^{n-1}$. We emphasize the interplay between $[{\bf DDW}]$ theory and the more general Lepagean theories. Indeed the  $[{\bf DDW}]$ manifold is not pataplectic whereas any open
subset of the universal multisymplectic manifold  $\Lambda^{n} T^{\star} {\mathfrak{Z}}$ necessary is.
 
 \
 
$\quad \mathfrak{3})$ {\bf Graded standpoint and copolarization}. Notice that if we focus on the {\em symmetry} point of view, then in contrast to the   ($n-1$)-form case where we have a natural link between Poisson brackets and dynamics, in the case of forms of arbitrary degrees the link is not clear. More precisely, $1 \leq p \leq n $ the $(n+1-p)$ multivector field $\xi_F$ defined such that $\hbox{d}F + \xi_F \iN \Omega = 0$ is not unique. It leads to a generalization of the Poisson bracket to the setting of graded structure \cite{Kana-01} \cite{Pau} \cite{Tul} in which it concentrates  on generalization of the Lie, Schouten-Nijenhuis and Fr\"{o}licher-Nijenhuis brackets.\footnote{In those works we build bracket operations on Hamiltonian multivector fields and Hamiltonian forms and the structure that emerges is a graded Poisson structure  built upon higher-order generalization of Gerstenhaber algebra.} However the lack of good dynamical properties for such forms suggests that we should give much more considerations to the notion of  $[{\bf OF}]$. In such a context, working on the basis  of the Einsteinnian picture, reflection on the very nature of observable forms led  F. H\'elein  and J. Kouneiher \cite{HK-02} \cite{HK-03} to the notion of {\em copolar forms}. The notion of copolarisation  allows us to define observable forms of any degree {\em collectively}   and hence leads beyond the setting of $[{\bf GR}]$.\footnote{In particular, it is shown in \cite{HK-02} that the notion of copolarization, apply in the context of Maxwell theory is completely coherent with physical phenomena} 
 
 \
 
$\quad \mathfrak{4})$ {\bf Slices and dynamical observables}.
We define a {\em slice} of codimension 1 as a submanifold $\Sigma \subset {\cal M}$ such that $T{\cal M} / T\Sigma$ is oriented and such that for any $\Gamma \in {\cal E}^{\cal H}$, $\Sigma$ is transverse to $\Gamma$. Also, let $\mathfrak{F} = \int_\Sigma F : {\cal E}^{\cal H} \longrightarrow \Bbb{R}$ be the observable
functional defined on the set of $n$-dimensional submanifolds ${\cal E}^{\cal H}$ as by the map : 
$ \Gamma \longrightarrow \mathfrak{F} (\Gamma) = {\int_{\Sigma\cap \Gamma} F}.
$ 
Then for any $F,G \in \mathfrak{P}_{\circ}^{n-1}$ the Poisson bracket (which coincides with the standard bracket on fields functionals) between two observables $\int_\Sigma F$ and $\int_\Sigma G$ is defined such that $\forall \Gamma \in {\cal E}^{\cal H}$ we have \eqref{PP}$(\mathfrak{i})$.
\begin{equation}\label{PP}
(\mathfrak{i}) \quad \left\{\int_\Sigma F,\int_\Sigma G\right\} (\Gamma) := \int_{ \Sigma \cap \Gamma} \{F,G\} 
 \quad
 \quad
 \quad
(\mathfrak{ii}) \quad  \left\{\int_{\Sigma^\circ} F,\int_{\Sigma^\bullet}  G\right\} (\Gamma) := \int_{\Sigma^{\circ} \cap \Gamma} \{F,G\} 
\end{equation}
Within the perspective of a  covariant theory, one would like to define a bracket  over two different slices $\Sigma^\circ $ and $ {\Sigma^{\bullet}} $. Given, $F,G \in \mathfrak{P}_{\circ}^{n-1}$ we then make  use of \eqref{PP}$(\mathfrak{ii})$ with the idea of {\em dynamical observables}. The notion of dynamical observable is formulated for $F \in \mathfrak{P}_{\circ}^{n-1}$ with the additional condition $\hbox{d} {\cal H}( { \xi_{F} }) = 0$.
 We denote by $ {\pmb{\cal O}}^{\cal H} $ the set of dynamical observables. An alternative description is made using the Poisson pseudobracket \eqref{gigi} and we find  suitable generalization of Dirac observable  formula \eqref{Dirac0}. $\forall F \in \mathfrak{P}^{n-1}$ an observable form is called dynamical (hence lives in $ {\pmb{\cal O}}^{\cal H} $) if $\{  {\cal H} , F \} = 0$. This condition is the sign of a homological feature :  if $\Gamma$ is a Hamiltonian $n$-curve, then this functional $\mathfrak{F} (\Gamma) $ depends only on the homology class of $\Sigma$. \cite{HK-02} \cite{HK-03} 

\

 $\quad \mathfrak{5})$ {\bf [CPS] and slices} (see \cite{H-02} for details) Let us sketch the outcome in the case of  $[{\bf CPS}]$
Let $\Gamma$ be an $n$-dimensional submanifold and let $\Sigma^{\circ}$ and 
$\Sigma^\bullet$ be two slices (hypersurfaces) in the same homology class. Finally, let $\mathcal{D} =\Sigma^{\bullet} - \Sigma^{\circ}$ be the portion of
$\mathcal{M}$ between $\Sigma^\circ$ and $\Sigma^\bullet$. Then, if $\delta\Gamma = \int_\Gamma\xi$ with $\xi$ a Jacobi field\footnote{A Jacobi vector field $\xi$ is a tangent vector field along $\Gamma$   solution of:
$ \forall \zeta \in C^{\infty} ({\cal M} , T_m{\cal M}),  ( \zeta \iN {\mathfrak{L}_\xi\Omega) |_\Gamma = 0}.$
}  and ${S^{\Sigma^\bullet}_{\Sigma^\circ}(\Gamma):= \int_{\Gamma\cap \mathcal{D}}\theta}$ is the action between $\Sigma^\circ\hbox{ and }
\Sigma^\bullet.$ We obtain the important formula \eqref{cpss} which crystalizes two related notions : the canonical pre-symplectic form ${\pmb{\varpi}}$ and the integral invariants theory. \cite{cartan}
\begin{equation}\label{cpss}
\underbrace{
{\delta (S^{\Sigma^\bullet}_{\Sigma^\circ})_\Gamma(\delta\Gamma) - \Theta^{\Sigma^\bullet}_\Gamma(\delta\Gamma)
+ \Theta^{\Sigma^\circ}_\Gamma(\delta\Gamma)}  }_{\bf{I}} =  \underbrace{ {\int_{\Gamma\cap \mathcal{D}} \xi \iN  \Omega } }_{\bf {II}} \quad \quad \hbox{with} \quad
{\delta \Gamma = \int_\Gamma\xi}
\quad \hbox{and}\quad
 {\Theta^\Sigma_\Gamma(\delta\Gamma):= \int_{\Gamma\cap\Sigma} \xi \iN \theta }
\end{equation}
 The term $[{\bf II}]$  in \eqref{cpss} vanishes if and only if $\Gamma$ is a Hamiltonian $n$-curve. Since ${ \Theta^{\Sigma^\bullet}
- \Theta^{\Sigma^\circ} = \delta (S^{\Sigma^\bullet}_{\Sigma^\circ})} $ is exact on $\mathcal{E}^{\cal H}$ then its differential  $\pmb{\varpi} = \delta\Theta^\Sigma$
does not depend on $\Sigma$ and is the canonical pre-symplectic  2-form on $\mathcal{E}^{\cal H}$ of $[{\bf CPS}]$ approach. Finally one observes the connection between $[{\bf MG}]$ and $[{\bf CPS}]$ from the point of view of Poincar\'e-Cartan  invariant integrals \cite{cartan} in the spirit of  the 
De Donder generalization. If $[{\bf I}]$ in \eqref{cpss}  vanishes for all $\Sigma^\bullet$ and $\Sigma^\circ$ and for all Jacobi fields $\xi$, then $\Gamma$ is a solution of the Hamilton equations. 

\

$\quad \mathfrak{6})$ {\bf Noether theorem} Some related work on the Noether theorem for covariant field theory is found in \cite{Fat} \cite{Forger}. In our context of $[{\bf MG}]$, and through the present discussion on observable
forms, the Noether theorem would be expressed as a correspondence between the set of dynamical
observable forms and the set of symmetries of the problem. Let us emphasize the main point : when
one works with the set $ {\pmb{\cal O}}^{\cal H} $ of observable $(n-1)$-forms, then one has a conservation law for any $ F \in  {\pmb{\cal O}}^{\cal H} $ along with a Hamiltonian $n$-curve : The Noether theorem simply gives $\hbox{d} F |_\Gamma = 0 $.

\

\section{The road ahead}

Let us conclude with some remarks aimed in two directions : Relativity and Quantization.

\
 
$\bullet$ {\bf Relativity Principle and Observables}. The notion of dynamical observable is related to a good definition of the Poisson bracket, (pseudobracket) which enforces good dynamical properties. Dynamics is seen in this picture as a matter of how to compare two observations, as made clear by the Relativity Principle. Notice also that in this approach Relativity is naturally expressed through the notion of copolarization. 

\

$\bullet$ {\bf Quantization}. We would like to give a quantum picture for $[{\bf MG}]$ - some attempt is made in \cite{Kana-02} \cite{H-02}.  In pursuing the search for $[{\bf QG}]$ we are faced with two related issues  : {\em non linearity} and {\em quantization}. The interplay of these two notions has been revealed in particular by H. Goldschmidt and S. Sternberg \cite{HGSS}, J. Kijowski \cite{JK-01}, F. H\'elein and J. Kouneiher, \cite{HK-02} \cite{FH-01}. Their works show that the two questions may be related. Let us cite F. H\'elein \cite{FH-01} {\em ''the quantization procedure works when the classical equation is linear but fails as
soon as the problem become nonlinear (interacting fields in the language of physicists)''.}  Since some part of the obstruction for
quantization can be understood as the fact that we don't get enough dynamical observables when the theory is  {\em non linear}, the presence of gauge symmetry can helps us to construct many more
dynamical observables. Actually, beside the gauge sector, one also overcomes these difficulties and finds
observable functionals with the tools of perturbations theory. This is the idea developed in \cite{FH-01}, and  applied by R. D. Harrivel in the study of interacting Klein-Gordon theory.\cite{RD-01}


\

{\bf Acknowledgments}

{\footnotesize{
{\em I am particularly grateful to Fr\'ed\'eric H\'elein and Joseph Kouneiher for invaluable discussions and  for having taught me so much about the topic of multisymplectic geometry. Also, for discussions of covariant canonical quantization and approaches to $[{\bf QG}]$, I address many thanks to John Stachel \cite{stachel02} for his communications on the sources and consequences of general covariance and background independence. Finally I am
particularly grateful to Michael Wright for his invaluable help in putting my text into  more fluent english.}
}}

\protect\label{conclusion}

\protect\label{mvf}
\pdfbookmark[1]{References}{ref}

\LastPageEnding

\end{document}